\documentstyle[12pt]{article}

\title{Relativistic Static Thin Disks with Radial Stress Support}
\author{Guillermo A. Gonz\'{a}lez\thanks{Permanent address: Escuela de
F\'{\i}sica, Universidad Industrial de Santander, A.A. 678, Bucaramanga,
Colombia, e-mail: guillego@uis.edu.co} \ and \ Patricio S.
Letelier\thanks{e-mail: letelier@ime.unicamp.br} \\ \\ Departamento de
Matem\'{a}tica Aplicada - IMECC \\ Universidade Estadual de Campinas \\
13081-970, Campinas, S.P., Brazil}

\date{ }

\begin{document}

\maketitle

\begin{abstract}
New solutions for static non-rotating thin disks of finite radius with
nonzero radial stress are studied. A method to introduce either radial
pressure or radial tension is presented. The method is based on the use
of conformal transformations.

\end{abstract}


\section{Introduction}

Axially symmetric solutions of Einstein field equations corresponding to
disklike configurations of matter are of great astrophysical interest
and have been extensively studied. These solutions can be static or
stationary and with or without radial pressure. Solutions for static
disks wi\-thout radial pressure were first studied by Bonnor and Sackfield
\cite{BOSA}, and Morgan and Morgan \cite{MM1}, and with radial pressure
by Morgan and Morgan \cite{MM2}, and, in connection with gravitational
collapse, by Chamorro, Gregory and Stewart \cite{CHGS}. Static disks with
radial tension may also be of interest and, although their relevance
in astrophysics is limited, these solutions can be studied as possible
sources of exact solutions for axially symmetric static problems.

Static thin disks without radial stress generate a Weyl spacetime. Then
they are described by two different metric functions \cite{WE1,WE2}. The
stability of these models can be explained by either assuming the
existence of hoop stresses or that the particles on the disk plane
move under the action of their own gravitational field in such a
way that as many particles move clockwise as counterclockwise. This
last interpretation is frequently made since it can be invoked to
mimic true rotational effects. A large class of statics thin disks
solutions were obtained by Letelier and Oliveira \cite{LO} using the
inverse scattering method. Solutions for self-similar static disks were
analyzed by Lynden-Bell and Pineault \cite{LP}, and Lemos \cite{LEM}. The
superposition of static disks with black holes were considered by
Lemos and Letelier \cite{LL1,LL2,LL3}, and Klein \cite{KLE}. Recently
Bi\u{c}\'{a}k, Lynden-Bell and Katz \cite{BLK} studied static disks as
sources of known vacuum spacetimes and Bi\u{c}\'{a}k, Lynden-Bell and
Pichon \cite{BLP} found an infinity number of new static solutions.

Static disks with radial stress are described by three metric functions so
presenting some degree of complexity and, despite its clear relevance, has
been seldom treated in the literature. The fact that the additional metric
function can be considered as the real part of an analytic function will
be the main ingredient of our method to introduce radial stress. Static
disks with radial stress can also be obtained in other ways. Lemos
\cite{LEM} studied self-similar static disks working in spherical polar
coordinates and using self-similarity to reduce the Einstein field
equations to an ordinary differential equation. Recently, Pichon and
Lynden-Bell \cite{PL} obtained static disks by making a cut through a
Weyl solution above all singularities or sources. The identification of
this solution with its mirror image yields disks with radial stress.

In this work we consider static non-rotating thin disks of finite radius
with nonzero radial stress. To obtain a well behaved radial stress we find
an analytical function compatible with the boundary conditions imposed
on the additional metric function by the requirement that the solution
be singularity-free outside the disk. We find a general expression
for the radial stress that can be written in terms of a square root in
such away that by choosing a sign solutions with radial pressure and
solutions with radial tension can be obtained. When we take the positive
(negative) sign of the square root, the stress is positive (negative)
and so we will have disks solutions with radial pressure (tension).

The plan of the paper is as follows. In Sec. 2 we present the principal
aspects of static thin disks solutions with radial stress, the expression
for the energy-momentum tensor of the disk and the general solution
that leads to disks with radial pressure or with radial tension. In
Sec. 3 we study solutions with radial pressure. In Sec. 4 solutions with
radial tension are considered. We conclude, in Sec. 5, by summaring our
main results.

\section{Static Thin Disks Solutions}

\subsection{The General Solution}

The metric for a static axially symmetric spacetime can be written in
quasicylindrical coordinates $(r, \varphi, z)$ in the form \cite{WALD,KSM}
\begin{equation} ds^2 = - e^\Phi dt^2 + R^2 e^{-\Phi} d\varphi^2 +
e^{\Lambda - \Phi} (dr^2 + dz^2), \label{eq:met} \end{equation} where $R$,
$\Phi$, and $\Lambda$ are functions of $r$ and $z$ only. In the vacuum,
the Einstein equations are equivalent to
\begin{eqnarray}
	& R,_{rr} + \ R,_{zz} = 0,  \label{eq:lap} \\ &
	\nonumber \\ & (R\Phi,_r),_r + \ (R\Phi,_z),_z = 0, \\ &
	\nonumber \\ & (R,^2_r + R,^2_z)\Lambda,_r = \frac{1}{2}RR,_r
	\left( \Phi,^2_r - \Phi,^2_z \right) + RR,_z\Phi,_r\Phi,_z + \
	(R,^2_r + R,^2_z),_r, \\ &				 \nonumber
	\\ & (R,^2_r + R,^2_z)\Lambda,_z = \frac{1}{2}RR,_z \left(
	\Phi,^2_z - \Phi,^2_r \right) + RR,_r\Phi,_r\Phi,_z + \ (R,^2_r +
	R,^2_z),_z, \label{eq:eq}
\end{eqnarray}
where $(\ ),_a = \partial /{\partial x^a}$.

In order to obtain a solution of (\ref{eq:lap}) - (\ref{eq:eq})
representing a thin disk located at $z = 0$, $0 \leq r \leq 1$, we assume
that the metric functions $R$, $\Phi$, and $\Lambda$ are continuous across
the disk, but have discontinuous first derivatives in the direction
normal to the disk. Althoug the radius of the disk has been taken as
unity, a suitable re-scaling of $r$ can be made. We will comeback to this
point later. The reflectional symmetry of (\ref{eq:lap}) - (\ref{eq:eq})
with respect to the plane $z = 0$ allows us to assume that $R$, $\Phi$,
and $\Lambda$ are even functions of $z$. Hence, $R,_z$, $\Phi,_z$ and
$\Lambda,_z$ are odd functions of $z$. We shall require that they not
vanish on the surfaces $z = 0^{\pm}$.

We now solve the Laplace equation (\ref{eq:lap}) in the region $r \geq
0$, $z > 0$ with the boundary conditions
\begin{eqnarray}
	R(0,z) \ &= \ 0, \qquad z > 0,		\\ &	   \nonumber
	\\ R,_z(r,0) \ &= \ 0, \qquad r > 1,	   \\ &       \nonumber
	\\ R,_z(r,0) \ &\neq \ 0, \qquad r \leq 1.
\end{eqnarray}
Let $\omega = r\ +\ iz$. We can consider $R$ as the real part of
an analytical function $W(\omega) = R(r,z)\ +\ iZ(r,z)$. Due to the
discontinuous behavior of $R,_z$ across the disk, the Cauchy-Riemann
equations imply that $Z$ is discontinuous across the disk while its normal
derivative $Z,_z$ is continuous. Then, the invariance of reflection
in the plane of the disk allows us to assume $W(r, -z) = \overline{W}
(r,z)$. We also require that $W(\omega)$ be locally one-to-one in the
outer region of the disk.

A simple solution compatible with these conditions is
\begin{equation}
	W(\omega) = \omega \pm \alpha \sqrt{\omega^2 - 1}, \label{eq:wab}
\end{equation}
where $\alpha$ is a positive constant, and the $\pm$
means that we can take any of the two possible signs of the square
root. When $z \geq 0$ we obtain
\begin{eqnarray}
	&\! R(r,z) \! &= r  \pm
	 \alpha \left[\frac{(|r^2 - z^2 - 1|^2 + 4r^2z^2)^{\frac{1}{2}} +
	 (r^2 - z^2 - 1)}{2}\right]^{\frac{1}{2}}\!, \label{eq:rab} \\
	&	&	\nonumber	\\ &\! Z(r,z) \! &= z  \pm
	 \alpha \left[\frac{(|r^2 - z^2 - 1|^2 + 4r^2z^2)^{\frac{1}{2}} -
	 (r^2 - z^2 - 1)}{2}\right]^{\frac{1}{2}}\!. \label{eq:zab}
\end{eqnarray}
For $z \leq 0$ we only need to replace $z \to - z$ in (\ref{eq:rab}) and
 (\ref{eq:zab}).

In order to solve the other field equations, we make a conformal
transformation
\begin{equation}
	r \rightarrow R(r,z), \qquad z \rightarrow Z(r,z), \label{eq:com1}
\end{equation}
in such a way that
\begin{equation}
	\Phi \rightarrow \tilde{\Phi} = \Phi, \qquad \Lambda \rightarrow
	\tilde{\Lambda} = \Lambda - ln|W'|^2, \label{eq:com2}
\end{equation}
where $W' = dW/d\omega$. With this transformation the
field equations take the usual Weyl form \cite{WE1,WE2}
\begin{eqnarray}
	&\tilde{\Phi},_{RR} \ + \ \frac{1}{R}\tilde{\Phi},_R \ + \
	\tilde{\Phi},_{ZZ} = 0, \label{eq:w1} \\ &		 \nonumber
	\\ &\tilde{\Lambda}[\tilde{\Phi}] = \frac{1}{2} \int R \ [ (
	\tilde{\Phi},^2_R - \tilde{\Phi},^2_Z ) dR \ + \ 2 \tilde{\Phi},_R
	\tilde{\Phi},_Z dZ ]. \label{eq:w2}
\end{eqnarray}
Note that (\ref{eq:w1}) is the Laplace equation in flat
three-dimensional space and so $\tilde{\Phi}$ can be taken as a solution
of Laplace equation for an appropriated Newtonian source with axial
symmetry. Once a  solution $\tilde{\Phi}$ is known, $\tilde{\Lambda}$
is computed from (\ref{eq:w2}) and so we obtain, from (\ref{eq:com1})
- (\ref{eq:com2}), a solution of the field equations (\ref{eq:lap}) -
(\ref{eq:eq}).

By using (\ref{eq:com2}) - (\ref{eq:w2}) we find the nonzero components
of the Riemann curvature tensor,
\begin{eqnarray}
	&R_{r\varphi r\varphi} &= \ \frac{R^2 R,_r}{2e^{\tilde{\Phi}}}
	[R,_r (\tilde{\Phi},_Z^2 \ - \ \tilde{\Phi},_R^2) + 2 R,_z
	\tilde{\Phi},_R \tilde{\Phi},_Z ],	 \label{eq:rie1}
	\\ &			   &	   \nonumber	   \\ &
	R_{r\varphi z\varphi} &= \ \frac{R^2 R,_r} {2e^{\tilde{\Phi}}}
	[R,_z (\tilde{\Phi},_Z^2 \ - \ \tilde{\Phi},_R^2) - 2 R,_r
	\tilde{\Phi},_R \tilde{\Phi},_Z ],	 \label{eq:rie2} \\
	&			&	\nonumber	\\ & R_{z\varphi
	z\varphi} &= \ R_{r\varphi z\varphi}, \label{eq:rie3}
\end{eqnarray}
and the Kretschmann scalar
\begin{eqnarray} &{R^{ab}}_{cd}
{R^{cd}}_{ab} &= \frac{e^{2(\tilde{\Phi} - \tilde{\Lambda})}}{R,_r^2 +
	R,_z^2} \times [ R,_z^2(\tilde{\Phi},_R^2 \ - \
	\tilde{\Phi},_Z^2)^2 \nonumber \\
&	&	\nonumber		\\ &	   &+ \ R,_r^2
(\tilde{\Phi},_R^4 \ + \
	6\tilde{\Phi},_R^2\tilde{\Phi},_Z^2 \ + \ \tilde{\Phi},_Z^4)
	\label{eq:kre} \\
&	&	\nonumber		\\ &	   &+ \ 4 R,_r R,_z
\tilde{\Phi},_R \tilde{\Phi},_Z(\tilde{\Phi},_R^2 \ -
	\ \tilde{\Phi},_Z^2) ]. \nonumber
\end{eqnarray}
From these expressions, using (\ref{eq:wab}) and the
properties of $\tilde{\Phi}$, we can test if the solutions have strong
curvature singularities.

\subsection{The Energy-Momentum Tensor}

Due to the discontinuous behavior of the derivatives of the metric
tensor across the disk, the Riemann curvature tensor contain Dirac
delta functions. The energy-momentum tensor can be obtained by the
distributional approach due to Papapetrou and Hamouni \cite{PH},
Lichnerowicz \cite{LICH}, and Taub \cite{TAUB}. See also Israel
\cite{IS1,IS2}. It can be written as ${T^a}_b = [{T^a}_b] \ \delta(z)$,
where $\delta$ is the Dirac function with support on the disk and
$[{T^a}_b]$ is the distributional energy-momentum tensor, which yield the
volume energy density and the principal stresses. The ``true'' surface
energy-momentum tensor of the disk can be written as ${\tau^a}_b =
e^{(\Lambda -\Phi)/2} \ [{T^a}_b]$.

The discontinuities in the first derivatives of the metric tensor can
be cast as \cite{LL2,LW}
\begin{equation} b_{ab} \ = \ g_{ab,z}|_{_{z =
0^+}} \ - \ g_{ab,z}|_{_{z = 0^-}}, 
\end{equation}
then the distributional
energy-momentum tensor is given by \begin{equation}
	[{T^a}_b] = \frac{1}{2}\{b^{az}{\delta^z}_b - b^{zz}{\delta^a}_b
	+ g^{az}{b^z}_b - g^{zz}{b^a}_b + {b^c}_c (g^{zz}{\delta^a}_b -
	g^{az}{\delta^z}_b)\}.
\end{equation}
For the metric (\ref{eq:met}) we obtain $[{T^a}_b] \equiv
diag (-\epsilon, \ p_{\varphi}, \ p_{r}, \ 0)$, with \begin{eqnarray}
	&\epsilon &= \ e^{\Phi - \Lambda} \left[ 2\Phi,_z -
	\Lambda,_z - \frac{2R,_z}{R} \right], \label{eq:emt1} \\ &
	&	\nonumber \\ &p_{\varphi} &= \ e^{\Phi - \Lambda} \
	[ \Lambda,_z ], \label{eq:emt2} \\ &	   &	   \nonumber \\
	&p_{r} &= \ e^{\Phi - \Lambda} \left[ \frac{2 R,_z}{R} \right],
	\label{eq:emt3}
\end{eqnarray}
where all these quantities are evaluated at $z = 0^+$, $0
\leq r \leq 1$. In order to obtain expressions for a disk with non-unit
radius, we only need to make the transformation $r \rightarrow a r$,
where $a$ is the radius of the disk.

From (\ref{eq:emt3}) and (\ref{eq:wab}) we can compute the general
expression for the radial stress $p_{r}$,
\begin{equation}
	p_{r} = \pm \ \frac{ 2 \alpha \sqrt{1 - r^2} }{1 + (\alpha^2 -
	1) r^2} \ e^{\tilde{\Phi} - \tilde{\Lambda}}, \label{eq:prr}
\end{equation}
where the $\pm$ correspond to the two signs of the square
root in (\ref{eq:wab}). Thus, when we take the positive (negative)
sign the stress is positive (negative), we have disks solutions
with radial pressure (tension). From the fact that $\tilde{\Phi}$ and
$\tilde{\Lambda}$ are both real and finite at the surface of the shells,
we can see that $p_{r}$ is a well behaved function of $r$ for any value
of $\alpha$.

By doing $z = 0$, $0 \leq r \leq 1$ in (\ref{eq:rab}) and (\ref{eq:zab})
we obtain the image of the disks by the conformal mapping,
\begin{equation}
	\alpha^2 R^2 \ + \ Z^2 \ = \ \alpha^2,
\end{equation}
so that the disks are mapped into spheroidal thin shells
of matter. According with this, the solution (\ref{eq:wab}) leads to
three different boundary conditions for a three-dimensional potential
problem. These boundary conditions correspond to the three possibilities
for $\alpha$: $\alpha = 1$, a spherical shell, $\alpha	> 1$, a prolate
spheroidal shell, and $0 < \alpha < 1$, an oblate spheroidal shell.

Using the above shells as sources to solve the Laplace equation
(\ref{eq:w1}) we obtain many different solutions of (\ref{eq:lap}) -
(\ref{eq:eq}). In all the cases the solutions of (\ref{eq:w1}) are
known and can be written in terms of Legendre functions. Due to the
discontinuous character of the Weyl coordinate $Z$ across the disk, a
polynomial in odd powers of $Z$ is discontinuous across the disk while
that a polynomial in even powers is a continuous function everywhere
but with discontinuous first derivatives with respect to $z$ across the
disk. Accordingly, the solutions of (\ref{eq:w1}) must be chosen with
the appropriated Legendre functions of even order in such away that the
solutions of (\ref{eq:lap}) - (\ref{eq:eq}) have the assumed behavior
at the disks.

In order to have a one-to-one correspondence in the outer region to
the disks, we need to choose the appropriate sign of the square root
in (\ref{eq:wab}) that map this region completely into the exterior or
interior region to the shells. The appropriate choose will depend of the
value of $\alpha$. When $\alpha \neq 1$ we only can take the positive sign
of the square root because with the negative sign the map (\ref{eq:wab})
is not single valued at the exterior of the disk and, in this case,
we only can take the solutions of (\ref{eq:w1}) that correspond to the
potential in the exterior of the appropriated spheroidal shell.

On the other hand, when $\alpha = 1$ we can take separately both of
the possibilities in (\ref{eq:wab}) and, for each one, we will obtain a
well posed boundary value problem. When we take the positive sign, the
outer region to the disk is mapped into the outer region to a spherical
shell of radius one. Thus we must take the solutions of (\ref{eq:w1})
that correspond to the potential in the exterior of the shell. When we
take the negative sign, the outer region to the disk is mapped into
the interior of the same shell and so the solutions of (\ref{eq:w1})
must be those that correspond to the interior potential.

\section{Static Disks with Radial Pressure}

For the positive sign of the square root in (\ref{eq:wab}) there are three
different families of solutions corresponding to the three possibilities
$\alpha = 1$, $\alpha < 1$ and $\alpha > 1$. In each case we consider
simple known Weyl solutions and from them, through (\ref{eq:com1}) and
(\ref{eq:com2}), we obtain the corresponding thin disks solutions. As we
shall see, in all the solutions considered we obtain disks with a well
behaved central sector where the energy-momentum tensor agree with the
weak and strong energy conditions \cite{HE}, but with a border region
where $\epsilon < 0$ in violation of the weak energy condition.

\subsection{Chazy-Curzon Disks}

The first family of solutions is obtained by choosing $\alpha = 1$
and transforming the disk into a spherical shell of radius one and the
outer region of the disk into the outer region to the shell. To solve
the Laplace equation (\ref{eq:w1}) we introduce spherical coordinates
$(\cal R, \theta)$, related to the Weyl coordinates by
\begin{equation}
	R = {\cal R} \sin \theta, \qquad Z = {\cal R} \cos \theta,
	\label{eq:ces}
\end{equation}
with $0 \leq {\cal R} \leq \infty$ and $-\frac{\pi}{2}
\leq \theta \leq \frac{\pi}{2}$. The shell is at ${\cal R} = 1$.

The general solution for the Laplace equation (\ref{eq:w1})
in the exterior of the shell, continuous across the disk, can be
written as
\begin{equation} \tilde{\Phi} = - \sum_{n = 0}^{\infty}
\frac{2C_{2n} P_{2n}(\cos \theta)}{{\cal R}^{2n + 1}},
\end{equation}
where $C_{2n}$ are constants and $P_{2n}(\cos \theta)$ are the usual
Legendre polynomials. The simplest case is the Chazy-Curzon solution
\cite{CHA,CUR}
\begin{equation} \tilde{\Phi} = -\frac{2\gamma}{{\cal R}}
, \qquad \tilde{\Lambda} = -\frac{\gamma^2 \sin^2 \theta}{{\cal R}^2},
\label{eq:esf}
\end{equation}
where $\gamma$ is a real constant. Note that $\tilde{\Phi}$ and 
$\tilde{\Lambda}$ are well behaved continuous functions for ${\cal R} 
\geq 1$ and vanish at infinity.

The metric tensor in Weyl coordinates is given by
\begin{eqnarray}
&g_{tt} &= \ - \ exp \left\{ \frac{-2\gamma}{\sqrt{R^2 + Z^2}}
\right\}, \\ &	     &	     \nonumber	     \\ &g_{\varphi\varphi}
&= \ R^2 exp \left\{ \frac{2\gamma}{\sqrt{R^2 + Z^2}} \right\}, \\
&	&	\nonumber	\\ &g_{rr} &= \ |W'|^2 exp \left\{
\frac{2\gamma}{\sqrt{R^2 + Z^2}} - \frac{\gamma^2R^2}{(R^2 + Z^2)^2}
\right\}, \\ &	     &	     \nonumber	     \\ &g_{zz} &= \ g_{rr}.
\end{eqnarray}
Using (\ref{eq:rab}), (\ref{eq:zab}) and (\ref{eq:com2}),
we can obtain its expression in the original coordinates. Then, taking
$z = 0^+$ and using (\ref{eq:emt1}) - (\ref{eq:emt3}), (\ref{eq:rab})
and (\ref{eq:zab}), we obtain
\begin{eqnarray} &\epsilon &= \ p_{r} \ [ 2
(\gamma - 1) - \gamma^2 r^2 ], \label{eq:es1} \\ &	 &	 \nonumber
\\ &p_{\varphi} &= \ p_{r} \ [ 1 + \gamma^2 r^2 ], \label{eq:es2} \\
&	&	\nonumber \\ &p_{r} &= \ p_0 \ \sqrt{1 - r^2} \ exp \{
\gamma^2 r^2 \} , \label{eq:es3}
\end{eqnarray}
where $p_0$ is a positive constant and $0 \leq r \leq 1$.

In order to analyze the behavior of the above solution we consider the
weak and strong energy conditions \cite{HE}. One of the requirements of
the weak energy condition is that $\epsilon \geq 0$ everywhere. From
(\ref{eq:es1}) we see that this condition is equivalent to
\begin{equation}
	2(\gamma - 1) - \gamma^2 r^2 \geq 0, \qquad 0 \leq r \leq 1.
\end{equation}
This equation has one zero at
\begin{equation}
	{r_0}^2 = \frac{2(\gamma - 1)}{\gamma^2}
\end{equation}
and so, in order to have $\epsilon > 0$ for $r = 0$,
we must impose the condition $\gamma > 1$. The maximum value for $r_0$
is obtained for $\gamma = 2$ and is given by ${r_0}^2 = 1/2$.

The strong energy condition require that $\sigma \geq 0$, where $\sigma =
\epsilon + p_{\varphi} + p_{r}$ is the effective Newtonian density. From
(\ref{eq:es1}) - (\ref{eq:es3}) we see that $\sigma = 2 \gamma p_{r}$
and so is positive everywhere on the disk for all value of $\gamma >
0$. In Figure 1 we plot the functions $\sigma$, $\epsilon$, $p_{r}$
and $p_{\varphi}$ in the interval $0 \leq r \leq 1$ for this choice
of the parameters, in $p_0$ units. The disk have a central region that
satisfy the weak and strong energy conditions and a border region where
$\epsilon < 0$, in violation of the weak energy condition.

\subsection{Zipoy-Voorhees Disks}

The second family of solutions is obtained by choosing $\alpha > 1$
in (\ref{eq:wab}), transforming the disk into a prolate spheroidal
shell. To solve the Laplace equation (\ref{eq:w1}) we introduce prolate
spheroidal coordinates $(\xi,\eta)$, related to the Weyl coordinates
by
\begin{equation}
	R^2 = \kappa^2(\xi^2 - 1)(1 - \eta^2), \qquad Z = \kappa\xi\eta,
	\label{eq:cpr}
\end{equation}
where $\kappa = \sqrt{\alpha^2 - 1}$, $1 \leq \xi
\leq \infty$ and $0 \leq \eta \leq 1$. The shell is located at $\xi =
\alpha/{\sqrt{\alpha^2 - 1}} > 1$.

The general solution for the exterior of the shell, continuous across
the disk, can be written as \begin{equation} \tilde{\Phi} = - \sum_{n
= 0}^{\infty} 2C_{2n} P_{2n}(\eta) Q_{2n}(\xi), \end{equation} where
$Q_{2n}$ are the Legendre functions of second kind. The simplest case
is the Zipoy-Voorhees solution \cite{ZIP,VOO}, also known as the Weyl
$\gamma$-solution \cite{WE1,WE2}, given by
\begin{equation}
	\tilde{\Phi} = \gamma \ln {\frac{\xi \ - \ 1}{\xi \ + \ 1}},
	\quad \tilde{\Lambda} = \gamma^2 \ln {\frac{\xi^2  - \ 1}{\xi^2  -
	\ \eta^2}}, \label{eq:pro}
\end{equation}
where $\gamma$ is a real constant. We see that
$\tilde{\Phi}$ and $\tilde{\Lambda}$ are continuous on the disk and
vanish at infinity.

The metric tensor in the prolate spheroidal coordinates is given by
\begin{eqnarray}
&g_{tt}	 &= \ - \ \left[\frac{\xi \ - \ 1}{\xi \
+ \ 1} \right]^{\gamma}, \\ &		    &	    \nonumber	    \\
&g_{\varphi\varphi}	&= \ R^2 \left[\frac{\xi \ + \ 1}{\xi \ - \ 1}
	\right]^{\gamma}, \\
&		&	\nonumber	\\ &g_{rr} &= \ |W'|^2
\left[\frac{\xi^2 - 1}{\xi^2 - \eta^2} \right]^{\gamma^2}\left[\frac{\xi \
+ \ 1}{\xi \ - \ 1}
	\right]^{\gamma}, \\
&		&	\nonumber	\\ &g_{zz} &= \ g_{rr}.
\end{eqnarray}
Using (\ref{eq:rab}), (\ref{eq:zab}), (\ref{eq:cpr})
and (\ref{eq:com2}), we can   obtain expressions in the original
coordinates. Taking the values of the above expressions at the disk and
using (\ref{eq:emt1}) - (\ref{eq:emt3}) we compute the energy density
and pressures,
\begin{eqnarray}
	&\epsilon &= \ p_{r} \left[\frac{2\gamma\sqrt{\alpha^2 - 1}
	- \alpha}{\alpha} - \frac{1 + (\alpha^2 - 1) \gamma^2 r^2}{1
	+ (\alpha^2 - 1) r^2} \right], \label{eq:pr1} \\ &	 &
	\nonumber \\ &p_{\varphi} &= \ p_{r} \left[ \frac{1 + (\alpha^2 -
	1)\gamma^2 r^2} {1 + (\alpha^2 - 1) r^2} \right], \label{eq:pr2}
	\\ &	   &	   \nonumber \\ &p_{r} &= \ p_0 \ \sqrt{1 - r^2} \
	[ 1 + (\alpha^2 - 1) r^2 ]^{(\gamma^2 - 1)}, \label{eq:pr3}
\end{eqnarray}
where $p_0$ is a positive constant and $0 \leq r \leq 1$.

From (\ref{eq:pr1}) we obtain the conditions
\begin{equation}
	\beta = \frac{\gamma \sqrt{\alpha^2 - 1}}{\alpha} \geq 1,
\end{equation}
to have $\epsilon \geq 0$ at $r = 0$, and
\begin{equation}
	\alpha^2 \beta^2 - 2 \alpha^2 \beta + (\alpha^2 + 1) \leq 0,
	\label{eq:e1}
\end{equation}
in order to have $\epsilon \geq 0$ at $r = 1$. We see
that $\gamma$ and $\alpha$ can not be adjusted to have $\epsilon \geq
0$ at the border of the disk and so $\epsilon(r)$ have a zero at $r_0$
given by
\begin{equation}
	{r_0}^2 = \frac{2 \alpha^2 (\beta - 1)}{\beta^2 - (1 -
	\alpha^2)(2\beta - 1)}. \label{eq:r0}
\end{equation}

Now, to have $\epsilon(r_0) = 0$ with real values of $\gamma$ and
$\alpha$, we obtain the conditions
\begin{equation}
	\alpha \leq \frac{1 - r_0^2}{r_0^2}, \qquad 0 < r_0^2 <
	\frac{1}{2}.
\end{equation}
The value of $\beta$ is given by
\begin{equation}
	\beta = \frac{1}{\alpha^2 r_0^2} \left[ 1 + (\alpha^2 - 1) r_0^2
	\pm \sqrt{ (1 - r_0^2)^2 - \alpha^2 r_0^4} \right]. \label{eq:gam}
\end{equation}
Then, by choosing a value of $r_0 < 1/\sqrt{2}$ we obtain
values for $\alpha$, $\beta$ and $\gamma$ so that the solution satisfy
the weak and strong energy conditions in the central region of the disk,
for $r < r_0$.

By choosing ${r_0}^2 = 0.2$ we can take $\alpha = 4$, $\beta = 5/4$
and $\gamma = \sqrt{5/3}$ so that the radial pressure is given by
\begin{equation}
	p_{r} = p_0 \sqrt{1 - r^2} (1 + 15 r^2)^{\frac{2}{3}}.
\end{equation}
From (\ref{eq:pr1}) - (\ref{eq:pr3}) we can see that
$\sigma = 2 \beta p_{r} > 0$ so that the solution satisfy the weak and
strong energy conditions in the central region of the disk. In Figure 2
we plot the functions $\epsilon$, $p_{r}$, $p_{\varphi}$ and $\sigma$
in the interval $0 \leq r \leq 1$ for this choice of the parameters,
in $p_0$ units. As in the first solution, the disk have a central region
that satisfy the weak and strong energy conditions and a border region
where $\epsilon < 0$, in violation of the weak energy condition.

\subsection{Bonnor and Sackfield Disks}

Finally, the third family of solutions is obtained by choosing $0 < \alpha
< 1$ in (\ref{eq:wab}) and transforming the disk into a oblate spheroidal
shell. To solve the Laplace equation (\ref{eq:w1}) we introduce oblate
spheroidal coordinates $(\zeta,\eta)$, related to the Weyl coordinates
by
\begin{equation}
	R^2 = \kappa^2(\zeta^2 + 1)(1 - \eta^2), \qquad Z =
	\kappa\zeta\eta, \label{eq:cob}
\end{equation}
where $\kappa = \sqrt{1 - \alpha^2}$, $0 \leq \zeta \leq
\infty$ and $0 \leq \eta \leq 1$. The shell is at $\zeta = \alpha/\sqrt{1
- \alpha^2} > 0$.

The general solution for the exterior of the shell, continuous across
the disk, can be written as
\begin{equation}
\tilde{\Phi} = - \sum_{n =0}^{\infty} 2C_{2n} P_{2n}(\eta) q_{2n}(\zeta),
\end{equation}
where $q_{2n}(\zeta) = i^{2n + 1} Q_{2n}(i\zeta)$. The simplest case is
given by
\begin{equation}
	\tilde{\Phi} = - 2 \ \gamma \ cot^{-1} \zeta, \quad
	\tilde{\Lambda} = - \gamma^2 \ln {\frac {\zeta^2  + \ 1}{\zeta^2 +
	\ \eta^2}}, \label{eq:obl}
\end{equation}
where $\gamma$ is a real constant. As in the above
two solutions, $\tilde{\Phi}$ and $\tilde{\Lambda}$ are continuous on
the disk and vanish at infinity. The above solution was found by Zipoy
\cite{ZIP} and Voorhees \cite{VOO} and interpreted by Bonnor and Sackfield
\cite{BOSA} as the gravitational field of a pressureless disk. We shall
see that the method based in the conformal transformation not only leads
to a nonzero radial pressure, but also to a nonzero azimuthal pressure.

From the metric tensor in oblate spheroidal coordinates
\begin{eqnarray}
&g_{tt}		&= \ - exp \left\{- 2 \gamma cot^{-1} \zeta \right\}, \\ &
&	\nonumber	\\ &g_{\varphi\varphi}	   &= \ R^2 exp \left\{2
\gamma cot^{-1} \zeta  \right\}, \\ &		    &	    \nonumber
\\ &g_{rr} &= \ |W'|^2 \ \left[ \frac{\zeta^2 + \eta^2}{1 + \zeta^2}
		\right]^{\gamma^2} exp \left\{2 \gamma cot^{-1} \zeta
		\right\}, \\
&		&	\nonumber	\\ &g_{zz} &= \ g_{rr},
\end{eqnarray}
we can obtain, using (\ref{eq:rab}), (\ref{eq:zab}),
(\ref{eq:cob}) and (\ref{eq:com2}), its expression in the original
coordinates. Taking the values of the above expressions at the disk and
using (\ref{eq:emt1}) - (\ref{eq:emt3}) we compute the energy density
and pressures
\begin{eqnarray}
	&\epsilon &= \ p_{r} \left[ \frac{2\gamma\sqrt{1 - \alpha^2} -
	\alpha}{\alpha} - \frac{1 + (1 - \alpha^2) \gamma^2 r^2}{1 +
	(\alpha^2 - 1) r^2} \right], \label{eq:ob1} \\ &       &
	\nonumber \\ &p_{\varphi} &= \ p_{r} \left[ \frac{1 + (1 -
	\alpha^2) \gamma^2 r^2} {1 + (\alpha^2 - 1) r^2} \right],
	\label{eq:ob2} \\ &	  &	  \nonumber \\ &p_{r} &= \frac{p_0
	\ \sqrt{1 - r^2}}{[ 1 + (\alpha^2 - 1) r^2 ]^{(\alpha^2 + 1)}},
	\label{eq:ob3}
\end{eqnarray}
where $p_0$ is a positive constant and $0 \leq r \leq 1$.

From (\ref{eq:ob1}) we obtain the condition
\begin{equation}
	\beta = \frac{\gamma \sqrt{1 - \alpha^2}}{\alpha} \geq 1,
\end{equation}
to have $\epsilon \geq 0$ at $r = 0$. As in the second
family of solutions, the condition (\ref{eq:e1}) is imposed in order
to have $\epsilon \geq 0$ at $r = 1$. Again, $\gamma$ and $\alpha$
can not be adjusted to have $\epsilon \geq 0$ at the border of the disk
and so $\epsilon(r)$ have a zero at $r_0$ given by (\ref{eq:r0}). Now,
to have $\epsilon(r_0) = 0$ with real values of $\gamma$ and $\alpha$,
we obtain
\begin{equation}
\alpha \leq \frac{1 - r_0^2}{r_0^2}, \qquad \frac{1}{2} < r_0^2 < 1,
\end{equation}
so that the value of $\beta$ is again given by
(\ref{eq:gam}). Then, by choosing a value of $1/\sqrt{2} < r_0 < 1$
we obtain limits for $\gamma$ and $\alpha$ so that the solution satisfy
the weak and strong energy conditions in the central region of the disk,
$r < r_0$.

By choosing ${r_0}^2 = 0.8$ we can take $\alpha = 1/4$, $\beta = 5$
and $\gamma = \sqrt{5/3}$ so that the radial pressure is given by
\begin{equation}
	p_{r} = \ \frac{p_0 \sqrt{1 - r^2}} {(1 - \frac{15}{16}
	r^2)^{\frac{8}{3}}}.
\end{equation}
From (\ref{eq:pr1}) - (\ref{eq:pr3}) we can see that
$\sigma = 2 \beta p_{r} > 0$ and so again we have a solution that
satisfy the weak and strong energy conditions for $0 \leq r \leq r_0$. In
Figure 3 we plot the functions $\epsilon$, $p_{r}$, $p_{\varphi}$ and
$\sigma$ in the interval $0.85 \leq r \leq 1$ for this choice of the
parameters, in $p_0$ units. As in the above two solutions, the disk have
a central region that satisfy the weak and strong energy conditions and
a border region where $\epsilon < 0$, in violation of the weak energy
condition. However, in this case the central well behaved region can be
extended to practically the entire disk by an appropriate selection of
the parameters $\gamma$ and $\alpha$.

\section{Statics disks with Radial Tension}

We now find some solutions with radial tension by taking the negative
sign of the square root in (\ref{eq:wab}). There is only one family
of solutions, corresponding to the case $\alpha = 1$, obtained by
transforming the disk into a spherical shell of radius one and the outer
region of the disk into the inner region to the shell.

To solve the Laplace equation (\ref{eq:w1}) we again introduce the
spherical coordinates (\ref{eq:ces}). The general solution for the Laplace
equation (\ref{eq:w1}) in the interior of the shell, continuous across
the disk, can be written as
\begin{equation}
	\tilde{\Phi} = - \sum_{n = 0}^{\infty} 2C_{2n} {\cal R}^{2n}
	P_{2n}(\cos \theta), \label{eq:ten}
\end{equation}
where $C_{2n}$ are constants. The simplest case is obtained
by taking the first term of the series and is given by
\begin{equation}
	\tilde{\Phi} = - 2 \gamma, \qquad \tilde{\Lambda} = 0,
\end{equation}
where $\gamma$ is a real constant.

The metric tensor is given by \begin{eqnarray}
	&g_{tt}			&= \ - \ e^{- 2 \gamma},	\\
	&			&	\nonumber		\\
	&g_{\varphi\varphi}	&= \ R^2 \ e^{ 2 \gamma },	\\ &
	&	\nonumber		\\ &g_{rr}		   &=
	\ |W'|^2 \ e^{ 2 \gamma },   \\ &			&
	\nonumber		\\ &g_{rr}		   &= \ g_{zz},
\end{eqnarray} where $R(r,z)$ and $|W'(\omega)|$ are obtained from
(\ref{eq:rab}) and (\ref{eq:zab}).

From (\ref{eq:rie1}) and (\ref{eq:rie2}), we see that ${R^{ab}}_{cd}
\equiv 0$ and so we have a spacetime defect \cite{LW}. The energy density
and tensions are given by
\begin{equation}
	\epsilon \ = \ - \ 2 p_{\varphi} \ = \ - 2 p_{r} \ = \ 2 p_0 \
	\sqrt{1 - r^2},
\end{equation}
where $p_0$ is a positive constant and $0 \leq r \leq
1$. In this case we have $\sigma = 0$ and we can see that this solution
satisfy all the energy conditions everywhere on the disk. This disk can
be interpreted as being formed by cosmic strings located along the radial
direction interwoven with loops. Another disklike topological defect, an
infinite disk with zero radial stress, was presented in \cite{LL3}. The
energy density and the azimuthal pressure were given by $\epsilon \ =
\ p_{\varphi} \ = \ 1$, also in agreement with the energy conditions.

A second solution is obtained by taking the second term of the series in
(\ref{eq:ten}). We find
\begin{equation}
	\tilde{\Phi} = \gamma ( R^2 - 2 Z^2 ), \qquad \tilde{\Lambda}
	= \frac{\gamma^2 R^2}{2} ( R^2 - 8 Z^2 ),
\end{equation}
where $\gamma$ is a real constant. As in the other
solutions, $\tilde{\Phi}$ and $\tilde{\Lambda}$ are continuous across
the disk and vanish at infinity, where $R = Z = 0$.

The metric tensor in Weyl coordinates is given by
\begin{eqnarray}
&g_{tt}		&= \ - exp \left\{ \gamma R^2 - 2 \gamma Z^2 \right\}, \\
&		&	\nonumber	\\ &g_{\varphi\varphi}	   &= \
R^2 exp \left\{ 2 \gamma Z^2 - \gamma R^2 \right\}, \\ &	       &
\nonumber	\\ &g_{rr} &= \ |W'|^2 exp \left\{\frac{\gamma^2
R^4}{2} - 4 \gamma^2 R^2 Z^2 - \gamma R^2 + 2 \gamma Z^2\right\},
\\ &		   &	   \nonumber	   \\ &g_{zz} &= \ g_{rr}.
\end{eqnarray}
Expressions in the original coordinates are obtained using
(\ref{eq:rab}), (\ref{eq:zab}), and (\ref{eq:com2}).

Taking the values of the above expressions at the disk and using
(\ref{eq:emt1}) - (\ref{eq:emt3}) we compute the energy density and
pressures,
\begin{eqnarray}
	&\epsilon &= - p_{r} \left[9 \gamma^2 r^4 - (6 + 8 \gamma)
	\gamma r^2 + 2 (2 \gamma + 1) \right], \label{eq:te1} \\ &
	&	\nonumber \\ &p_{\varphi} &= - p_{r} \left[8 \gamma^2
	r^2 - 9 \gamma^2 r^4 - 1 \right], \label{eq:te2} \\ &	    &
	\nonumber \\ &p_{r} &= - p_0 \ exp \left\{(3 + 4 \gamma)\gamma
	r^2 - 4.5 \gamma^2 r^4\right\} \sqrt{1 - r^2}, \label{eq:te3}
\end{eqnarray}
where $p_0$ is a positive constant and $0 \leq r \leq 1$.

From (\ref{eq:te1}) we can see that the weak energy condition is
equivalent to
\begin{equation}
	9 \gamma^2 r^4 - (6 + 8 \gamma) \gamma r^2 + 2 (2 \gamma + 1)
	\geq 0, \label{eq:den}
\end{equation}
for $0 \leq r \leq 1$. This expression have two zeros at
\begin{equation}
	r_0^2 = \frac{3 + 4 \gamma \pm \sqrt{16 \gamma^2 - 12 \gamma -
	9}} { 9 \gamma}, \label{eq:r02}
\end{equation}
and so there are three different classes of solutions
corresponding to the values of $\gamma$ such that the above equation
have either two, one or any real positive root.

When $\gamma < - 0.5$ there is a solution $r_0 < 1$ of (\ref{eq:r02})
in such away that the disks have central regions with $\epsilon < 0$ and
border regions with $\epsilon > 0$. When $- 0.5 \leq \gamma \leq 3(1 +
\sqrt{5})/8$ we can see that $\epsilon \geq 0$ everywhere on the disks,
accordingly the weak energy condition is satisfied. Finally, when $\gamma
> 3(1 + \sqrt{5})/8$, there are two different real positive solutions of
(\ref{eq:r02}), $r_1$ and $r_2$, so that the disks have a central region
with $\epsilon > 0$, when $r < r_1$, a region with $\epsilon < 0$,
when $r_1 < r < r_2$, and a region with $\epsilon > 0$, when $r > r_2$.

From (\ref{eq:te1}) - (\ref{eq:te3}) we can see that
\begin{equation}
	\sigma = - \gamma p_r ( 4 - 6 r^2).
\end{equation}
Hence $\sigma = 0$ at $r^2 = 2/3$. Accordingly, the
solutions are in agree with the strong energy condition only in a
region, the central region of the disks when $\gamma > 0$, or the border
region of the disks when $\gamma < 0$. We also see from (\ref{eq:te2})
that the azimuthal stress is zero at \begin{equation} r_{\pm}^2 \ = \
\frac{4 \gamma \pm \sqrt{16 \gamma^2 - 9 }}{9 \gamma}.	\end{equation}
When $|\gamma| > 0.75$ there are two different positive roots, $r_- < r_+
< 1$. Hence there is a region of the disks where the azimuthal stress
is positive, for $r_- < r < r_+$. On the other hand, when $|\gamma| <
0.75$ the a\-zi\-muthal stress is negative everywhere.

In order to show the behavior of the different classes of solutions
that can be obtained we consider some particular values of $\gamma$ . In
Figure 4 -- 7 we plot the functions $\epsilon$, $p_{r}$, $p_{\varphi}$
and $\sigma$ in the interval $0 \leq r \leq 1$ for $\gamma = - 1$,
$\gamma = - 0.5$,  $\gamma = 1$ and $\gamma = 2$, respectively, in $p_0$
units. We see that very different classes of solutions can be obtained,
some of them with either the center of the disk or the border in accord
with the energy conditions.

\section{Concluding Remarks}

In this work a method was presented which can be used to obtain static
thin disks models with a well behaved nonzero radial stress. The general
expre\-ssion that was obtained for the radial stress admits solutions
with positive stress, representing disks with radial pressure support,
and solutions with negative stress, representing disks with radial tension
support. Is easy to shown that, with the simple solution (\ref{eq:wab}),
when we consider solutions with positive stress is not possible to fulfill
all the energy conditions at the entire disk, whereas that with negative
stress we can obtain solutions that satisfy all the energy conditions
everywhere on the disks, as the first solution studied at Sec. 4.

The method that was described can be used to obtain many static disks
solutions with a radial pressure support that presents a very reasonable
phy\-sical behavior and perhaps, by considering another more involved
solutions of the Laplace equation (\ref{eq:lap}), the solution can be
adjusted to fulfill the weak and strong energy conditions everywhere
on the disks, leading so to a physically acceptable source which would
produce the considered vacuum solutions. Also, as we will do in a future
paper, the method can be extended to consider rotating thin disks,
which have a stronger astrophysical interest.

\section*{Acknowledgments}

G.A.G. thanks a grant of CAPES - Brazil and a grant of COLCIENCIAS -
Colombia. P.S.L. thanks grants of CNPq and FAPESP - Brazil.

\end{document}